\documentclass{article}
\usepackage[utf8]{inputenc}
\usepackage{authblk}
\usepackage{graphicx}
\usepackage{multirow}
\usepackage{biblatex}
\title{Towards Super-Resolution CEST MRI for Visualization of Small Structures}

\author{
 Lukas Folle$^{1,*}$,
 Katharian Tkotz$^{2,*}$,
 Fasil Gadjimuradov$^1$,
 Lorenz Kapsner$^2$,
 Moritz Fabian$^3$,
 Sebastian Bickelhaupt$^2$,
 David Simon$^4$,
 Arnd Kleyer$^4$,
 Gerhard Krönke$^4$,
 Moritz Zaiß$^3$,
 Armin Nagel$^{2,5}$,
 Andreas Maier$^1$
}

\affil[1]{Pattern~Recognition~Lab, Friedrich-Alexander-Universit\"at~Erlangen-N\"urnberg~(FAU), Erlangen, Germany}
\affil[2]{Institute~of~Radiology, University~Hospital~Erlangen, Friedrich-Alexander-Universit\"at~Erlangen-N\"urnberg (FAU), Erlangen, Germany}
\affil[3]{Institute~of~Neuroradiology, University~Hospital~Erlangen, Friedrich-Alexander-Universit\"at~Erlangen-N\"urnberg (FAU), Erlangen, Germany}
\affil[4]{Department of Internal Medicine 3, University~Hospital~Erlangen, Friedrich-Alexander-Universit\"at~Erlangen-N\"urnberg (FAU), Erlangen, Germany}
\affil[5]{Division of Medical Physics in Radiology, German Cancer Research Center (DKFZ), Heidelberg, Germany}
\affil[*]{Equal contribution}

\begin{document}

\maketitle

\begin{abstract}
The onset of rheumatic diseases such as rheumatoid arthritis is typically subclinical, which results in challenging early detection of the disease. However, characteristic changes in the anatomy can be detected using imaging techniques such as MRI or CT. Modern imaging techniques such as chemical exchange saturation transfer (CEST) MRI drive the hope to improve early detection even further through the imaging of metabolites in the body. To image small structures in the joints of patients, typically one of the first regions where changes due to the disease occur, a high resolution for the CEST MR imaging is necessary. Currently, however, CEST MR suffers from an inherently low resolution due to the underlying physical constraints of the acquisition. In this work we compared established up-sampling techniques to neural network-based super-resolution approaches. We could show, that neural networks are able to learn the mapping from low-resolution to high-resolution unsaturated CEST images considerably better than present methods. On the test set a PSNR of 32.29\,dB (+10\%), a NRMSE of 0.14 (+28\%), and a SSIM of 0.85 (+15\%) could be achieved using a ResNet neural network, improving the baseline considerably.
This work paves the way for the prospective investigation of neural networks for super-resolution CEST MRI and, followingly, might lead to a earlier detection of the onset of rheumatic diseases.
\end{abstract}

\section{Introduction}
Patients affected by rheumatic diseases such as psoriatic arthritis (PsA) or rheumatoid arthritis (RA) suffer from synovial inflammation, which can lead to cartilage and bone destruction if no treatment is initiated~\cite{doi:10.1056/NEJMra1004965}. But, if detected at an early stage, rheumatic diseases can be controlled well through the prescription of disease modifying anti-rheumatic drugs (DMARD)~\cite{Schett1428}. The early detection, however, still remains a challenging task, even with the use of neural networks~\cite{Folle2021}. Chemical exchange saturation transfer (CEST) magnetic resonances imaging (MRI) of joints allows for the visualization of biochemical alterations of tissue and, thus, is a promising tool for the detection of disease onset. One major obstacle of CEST MRI is the inherently low resolution~\cite{Wu2016}. This is due to the fact that a high signal to noise ratio (SNR) is needed for a reproducible CEST contrast, which restricts image resolution. For the application in a rheumatic setting, imaging of thin structures such as the joint tissue of the knee is thus very challenging.

Super-resolution focuses on the task of computing high-resolution (HR) images from low-resolution (LR) images by exploiting prior information~\cite{kohler2016robust}. One very prominent method for extracting and exploiting this information are neural networks~\cite{kohler2019toward}. Previously, neural network-based super-resolution mostly focused on increasing the resolution of common MR sequences such as fast-spin-echo and double echo in steady-state~\cite{GREENSPAN2002437,Chaudhari}.
So far, no evaluation of super-resolution techniques for CEST MRI has been performed. One established method along different MRI sequences is k-space zero-filling and thus will serve as the baseline for our comparison~\cite{LUO20171}.

In this work, we will investigate the feasibility of applying super-resolution neural networks for CEST MRI scans by comparing different neural network architectures and loss functions for 7T unsaturated knee CEST MR images.

\begin{figure}[t]
    \begin{center}
        \makebox[\textwidth]{\includegraphics[width=0.55\paperwidth]{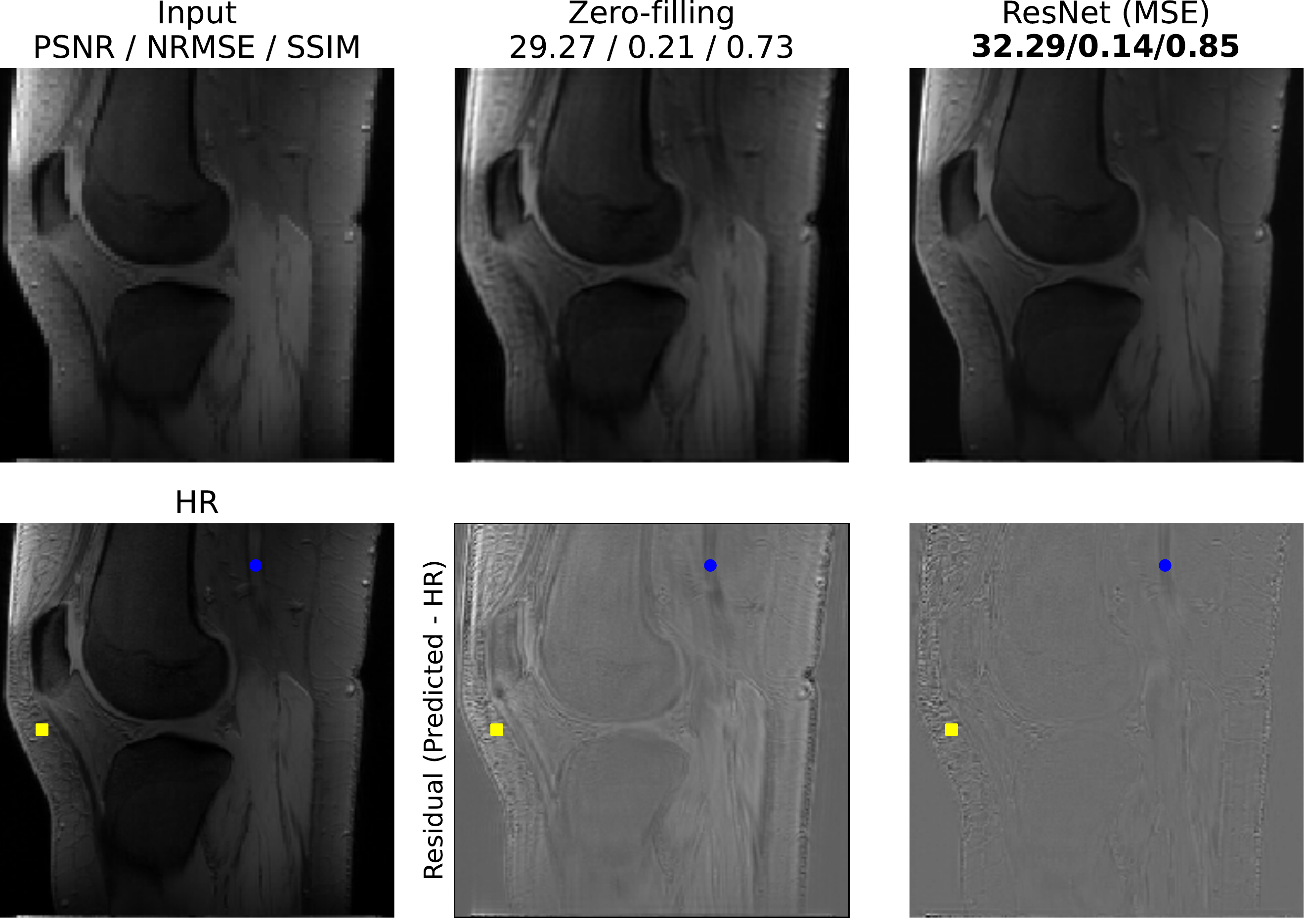}}
    \end{center}
    \caption{Comparison of traditional k-space zero-filling and the best performing neural network prediction. Input to both methods is the low-resolution image LR (top left) and target is the ground-truth high-resolution image HR (bottom left). Residuals between prediction and HR are depicted below the predictions (pixel values in [-0.23, 0.38] for all residuals). Yellow square and blue dot mark regions with high difficulty for both methods due to high-frequency image components or blood flow.}
    \label{fig:3}
\end{figure}

\section{Theory \& Methods}
In the following, the acquisition of the CEST MR scans will be described, followed by an overview of the deep learning methods and current interpolation techniques.

The main motivation of CEST MRI is the imaging of other metabolites than free water, which mainly contributes to normal 1H images - described in~\cite{https://doi.org/10.1002/mrm.22761,GUIVELSCHAREN199836}. This contrast utilizes the chemical exchange of bound hydrogen atoms with free water. To add metabolic information, the magnetization at the resonance frequency of the bound hydrogen proton is saturated. Due to the chemical exchange the saturation is transferred to the free water signal and results in a signal decrease in the subsequent acquired MR image. Typically, images with saturation at different resonance frequencies are acquired, resulting in voxel-wise spectral information showing the abundance of the metabolic groups. However, due to the repeated flipping and relaxation of the magnetization during the image acquisition the magnetization converges to a steady state, that is independent of the previous saturation. Thus, a higher resolution, which corresponds to a longer acquisition phase, will reduce the metabolic weighting in the resulting image. The only possibility to avoid this is a segmented k-space acquisition, which, however, prolongs the acquisition process significantly.

For the training of super-resolution networks, however, HR data is necessary. For our experiments, the training and test data consisted only of unsaturated images. All scans were acquired using a 7\,T MR Siemens Magnetom Terra (Siemens Healthcare, Erlangen, Germany) with a 1Tx/28Rx knee coil (Quality Electrodynamics LLC, Mayfield Village, Ohio, USA). The high-resolution CEST images were acquired with a 3D GRE sequence~\cite{https://doi.org/10.1002/nbm.3879} and had a slice thickness of 2.5\,mm and a pixel spacing of 0.802\,mm resulting in volume dimensions of $192 \times 192 \times 40$. A repetition time of 13ms was chosen, including three echos at 1.81\,ms, 6.43\,ms, and 11.05\,ms to achieve a higher variety of the image contrast.
In total, 39 HR volumes were available from 13 healthy volunteers with three different echo times that were split in to 23 training volumes, 8 validation, and 8 testing volumes. No data was shared across the three sets. All participants gave written informed consent before the measurement and were recruited under the approval of the local ethics committee~(Ethik-Kommission, Friedrich-Alexander-Universit\"at, Erlangen-N\"urnberg).

For the training and quantitative comparison, LR images were generated by down-sampling of the HR images by a factor of two. Trilinear interpolation as a further established method was dropped due to the inferior performance to the k-space zero-filling.

Pre-processing steps included z-score normalization and, for the training phase, random 3D crops of size ($64 \times 64 \times 16$), random flipping along all three axes, and random rotations (15 degree in-plane, 5 degree through-plane).
To estimate how close a predicted image is to the HR image, structural similarity (SSIM), peak signal to noise ratio (PSNR), and normalized root mean square error (NRMSE) were evaluated.

Two different network architectures were compared throughout this work, namely a ResNet-based network~\cite{7780459} and a DenseNet-based network~\cite{8099726}, both adapted to 3D images. A Wasserstein generative adversarial network (GAN)~\cite{10.5555/3305381.3305404} during initial experiments could not be trained to convergence and thus was removed from the comparison.
The DenseNet-like model consisted of four DenseBlocks with a growth rate in each block of 16 and 32 initial channels resulting in a total of 411k parameters. The ResNet-like model consisted of 16 residual blocks each with 32 channels leading to 528k parameters. For the ResNet and the DenseNet, different loss functions were compared: mean-squared error loss (MSE), SSIM loss, and a perceptual loss. The perceptual loss was calculated based on the distance of feature vectors extracted from a VGG network trained in ImageNet~\cite{Zhang2018TheUE,5206848}.
Training was stopped when no improvement on the validation set SSIM was perceived for 100 epochs. Adam with a constant learning rate of $1e^{-4}$ was used for all models. The complete dataset was split into 60\% training samples, 20\% validation samples, and 20\% test samples.

For comparison to the predictions of the neural networks, k-space zero-filling was used. K-space zero filling is applied to the raw k-space MR data. Thereby the outer regions of the k-space, which contain the high frequency information, are padded with zeros in all three directions up to the desired resolution. Typically, filters are applied to smooth the hard edges in k-space, but were removed in our comparison to achieve higher SSIM and PSNR.

\section{Results}
Table~\ref{tab:1} shows the quantitative comparison of the different neural network configurations together with the baseline, k-space zero-filling on the test set.
\begin{table}[t]
    \centering
    \begin{tabular}{ccccc}
    Method & \multicolumn{1}{l|}{Loss function} & PSNR (SD) $\uparrow$ & NRMSE (SD) $\downarrow$ & SSIM (SD) $\uparrow$ \\ \hline
    \multicolumn{1}{r}{K-space zero-filling} & \multicolumn{1}{c|}{n/a}        & 29.27 (0.77) & 0.21 (0.04) & 0.73 (0.03) \\ \hline
    \multirow{3}{*}{ResNet}                  & \multicolumn{1}{c|}{MSE}        & \textbf{32.29} (1.19) & \textbf{0.14} (0.03) & \textbf{0.85} (0.01) \\
                                             & \multicolumn{1}{c|}{SSIM}       & 31.08 (0.98) & 0.16 (0.03) & 0.84 (0.01) \\
                                             & \multicolumn{1}{c|}{Perceptual} & 31.65 (0.99) & 0.18 (0.04) & 0.77 (0.03) \\ \hline
    \multirow{3}{*}{DenseNet}                & \multicolumn{1}{c|}{MSE}        & 31.07 (0.87) & 0.16 (0.03) & 0.83 (0.01) \\
                                             & \multicolumn{1}{c|}{SSIM}       & 30.70 (1.02) & 0.17 (0.03) & 0.83 (0.01) \\
                                             & \multicolumn{1}{c|}{Perceptual} & 31.93 (1.10) & 0.17 (0.04) & 0.80 (0.02)
    \end{tabular}
    \caption{Comparison of k-space zero-filling and the neural network for the test cases of the dataset. PSNR: peak signal to noise ratio (dB); (NR) MSE: (normalized root) mean square error; SSIM: structural similarity; SD: standard deviation; $\uparrow$: higher is better; $\downarrow$: lower is better. Best performance marked in bold.}
    \label{tab:1}
\end{table}
Overall, the ResNet models outperformed the DenseNet models consistently across all metrics and loss functions except for the perceptual loss. Further, mean-squared error loss leads to better results for both models. Finally, k-space zero-filling could be considerably outperformed using the best neural network configuration by an increase of the performance using the ResNet trained with MSE loss of PSNR +3.02 dB, NRMSE +0.07, and SSIM +0.12.

Figure~\ref{fig:3} provides a qualitative comparison of the neural-network-based method (ResNet MSE loss) over the currently used method (k-space zero-filling) together with the LR input and the HR ground-truth (GT). It can be seen that the residual between the prediction and the GT is substantially improved by the network over the zero-filling approach. Regions the network is not able to recover typically have either very high frequencies such as subcutaneous fat anterior to the patella (yellow square) or strong blood flow such as the popliteal artery (blue dot). The regions of high interest for arthritis research, the tibiofemoral joint and patellofemoral joint, are resolved with high accuracy by the network.

\section{Discussion}
In this work, we demonstrated the application of neural network for super-resolution on CEST MR imaging and compared different network architectures and loss functions. We believe, that the results of this work have the potential to further advance the application of the CEST contrast in arthritis research, as our findings enable imaging of smaller structures in the knee joint which play a vital role in the early detection of rheumatic diseases. The presented work considerably outperformed the baseline method. Overall, ResNet inspired models as well as the mean-squared error loss function performed best in our experiments.

Our approach has some limitations. We did not test the combination of loss functions and the implication of the loss function on the image appearance. In future work, we want to focus on those points specifically.
Additionally, the general reasoning behind the predictions of neural networks is inherently hard to understand and errors in the prediction might be hard to spot. Thus, in future work, we want to focus on the data consistency of the predictions and provide an easy to use software tool to compare LR image and HR prediction at the click of a button. Further, we want to make use of all the information from the k-space by incorporating the reconstruction step into our method. Currently, parts of this information are lost due to the lossy reconstruction.
Finally, we aim to perform a qualitative evaluation of the best performing configuration on prospectively acquired low-resolution CEST MRI at different frequencies and compare the resulting CEST contrast with currently used methods.

\section*{Acknowledgements}
This work was supported by the emerging field initiative (project 4 Med 05 “MIRACLE”) of the University Erlangen-N\"urnberg and MASCARA - Molecular Assessment of Signatures Characterizing the Remission of Arthritis Grant 01EC1903A. We would like to thank the d.hip data center for computational support.

\printbibliography

\end{document}